\documentclass[
reprint,
superscriptaddress,
amsmath,
amssymb,
]{revtex4-2}

\usepackage{graphicx}
\usepackage{dcolumn}
\usepackage{bm}
\usepackage{hhline}
\usepackage[switch]{lineno}
\usepackage{array}
\usepackage{xcolor}
\usepackage{soul}
\usepackage[indent,skip=6pt]{parskip}

\begin{document}

\title{\textcolor{black}{High-Performance Near-Infrared Quantum Emission from Color Centers in hBN}}



\author{Sean Doan}
\altaffiliation{These authors contributed equally to this work.}
\affiliation{Physics Department, University of California, Santa Barbara, CA 93106, USA}

\author{Sahil D. Patel}
\altaffiliation{These authors contributed equally to this work.}
\affiliation{\mbox{Electrical and Computer Engineering Department, University of California, Santa Barbara, CA 93106, USA}}

\author{Yilin Chen}
\affiliation{\mbox{Materials Department, University of California, Santa Barbara, CA 93106, USA}}

\author{Jordan A. Gusdorff}
\affiliation{\mbox{Electrical and Systems Engineering Department, University of Pennsylvania, Philadelphia, PA, 19104, USA}}
\affiliation{\mbox{Materials Science and Engineering Department, University of Pennsylvania, Philadelphia, PA, 19104, USA}}

\author{Mark E. Turiansky}
\affiliation{\mbox{US Naval Research Laboratory, 4555 Overlook Ave SW, Washington, DC 20375, USA}}

\author{Luis Villagomez}
\affiliation{\mbox{Electrical and Computer Engineering Department, University of California, Santa Barbara, CA 93106, USA}}

\author{Luka Jevremovic}
\affiliation{\mbox{Electrical and Computer Engineering Department, University of California, Santa Barbara, CA 93106, USA}}

\author{Nicholas Lewis}
\affiliation{\mbox{Electrical and Computer Engineering Department, University of California, Santa Barbara, CA 93106, USA}}

\author{Kenji Watanabe}
\affiliation{Research Center for Functional Materials, National Institute for Materials Science, 1-1 Namiki, Tsukuba 305-004, Japan}

\author{Takashi Taniguchi}
\affiliation{International Center for Materials Nanoarchitectures, National Institute for Materials Science, 1-1 Namiki, Tsukuba 305-004, Japan}

\author{Lee C. Bassett}
\affiliation{\mbox{Electrical and Systems Engineering Department, University of Pennsylvania, Philadelphia, PA, 19104, USA}}

\author{Chris Van de Walle}
\affiliation{\mbox{Materials Department, University of California, Santa Barbara, CA 93106, USA}}

\author{Galan Moody}
\email{moody@ucsb.edu}
\affiliation{\mbox{Electrical and Computer Engineering Department, University of California, Santa Barbara, CA 93106, USA}}

\date{\today}


\begin{abstract}

\textcolor{black}{Color centers hosted in hexagonal boron nitride have emerged as a highly promising platform for single-photon emission and spin-photon technologies relevant to quantum communication and quantum networking. As a wide-bandgap van der Waals material, hBN can host optically active quantum defects across a broad spectral range. Here, we demonstrate a simple and scalable oxygen-plasma process that reproducibly creates single quantum emitters in hBN with blinking-free zero-phonon lines spanning the near-infrared from 700 up to  971~nm. These emitters combine MHz-level brightness, single-photon purity up to 99.9\%, and ultranarrow cryogenic linewidths down to 2.7~GHz under quasi-resonant excitation, placing them in a particularly attractive regime for quantum photonics. Photostability measurements further reveal resistance to photobleaching, sub-nm spectral stability over long timescales, and near-shot-noise-limited intensity fluctuations. Analysis of the phonon sidebands shows weak vibronic coupling and ZPL-dominated emission, with Debye--Waller factors approaching 50\%. Control experiments together with EDS elemental mapping support oxygen incorporation as a necessary ingredient in activating the NIR emitter population, while first-principles calculations identify O$_N$V$_N$ and O$_N$V$_N$H as the leading defect candidates. These results establish a high-performance NIR quantum-emitter platform in hBN for free-space quantum networking and future integrated quantum-photonic architectures.}


\end{abstract}

 

\maketitle
\vspace{-5pt}

\begin{figure*}[t!] \centering
     \includegraphics[scale=1.4]{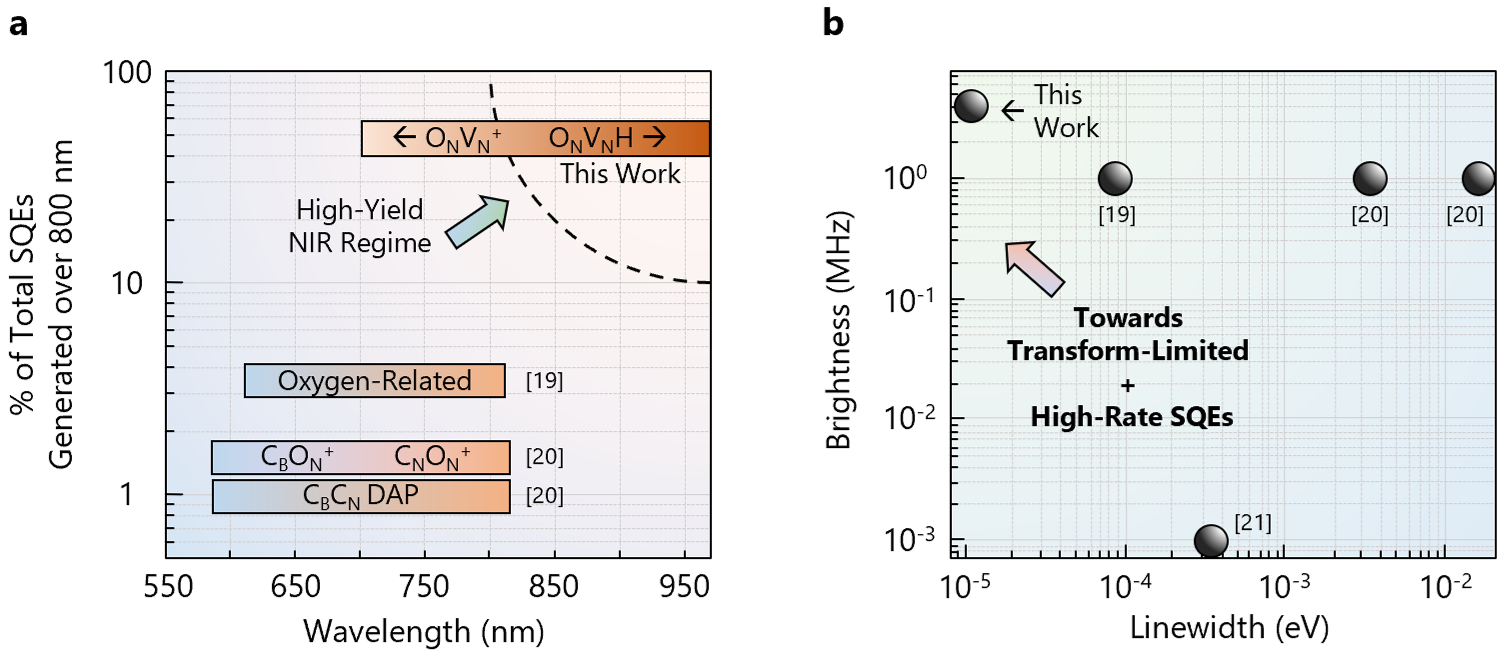}  
          \caption{\footnotesize \textbf{Literature benchmarking of the statistical prevalance and photophysical performance of NIR hBN SQEs.} \textcolor{black}{\textbf{(a)} Comparison of the fraction of total SQEs generated above 800 nm versus emission wavelength range for this work and representative prior reports on hBN SQEs relevant to oxygen- or plasma-related defect formation. The horizontal bars denote the reported spectral ranges associated with the corresponding defect families or emitter classes, while their vertical placement indicates the reported percentage of emitters lying in the $>$800 nm regime. The SQEs generated in this work occupy the high-yield NIR regime with a substantial prevalence above 800 nm. Within the \textit{This Work} box, the arrows indicate that the two proposed defect classes are distributed differently across this overall spectral band, with one defect (O$_N$V$_N$) exhibiting a comparatively bluer wavelength distribution and the other (O$_N$V$_N$H) exhibiting a comparatively redder wavelength distribution within the full bandwidth span of the observed emitters. \textbf{(b)} Comparison of brightness versus linewidth for NIR hBN SQE studies. The point corresponding to this work lies in the regime of simultaneously high brightness and narrow linewidth. The shaded arrow indicates the general direction toward transform-limited, high-rate SQEs, and the comparison points are taken from literature values summarized in the Supporting Information.}}
          \label{fig: novelty}
\end{figure*}

\textcolor{black}{Color centers in hexagonal boron nitride (hBN) have developed into a promising platform for single-photon generation and spin-optical functionalities in van der Waals materials \cite{ turunen2022quantum, azzam2021prospects}, owing to their chemical robustness \cite{cai2019high}, atomically thin geometry, and compatibility with photonic, optoelectronic \cite{li2021integration,parto2022cavity,yamashita2025deterministic,patel2024surface}, and heterostructure-based device architectures \cite{grosso2017tunable, luo2018deterministic, zhao2021site}. At the same time, the practical utility of hBN quantum emitters remains strongly tied to the specific defect family being addressed \cite{cassabois2016hexagonal,fournier2021position,gottscholl2020initialization}, since the optical properties reported to date span a wide range of linewidths, vibronic coupling strengths, spectral stability, and emission wavelengths. In particular, the search for high-quality emitters in the near-infrared (NIR) has become increasingly important because this spectral window is favorable for free-space quantum links, reduced atmospheric attenuation, and eventual compatibility with photonic networking technologies \cite{gottscholl2021room,robertson2023detection,gong2023coherent}. Yet, while hBN can host defect emission extending into the red and NIR, many previously studied emitters have remained concentrated in the visible, exhibited pronounced phonon sidebands, or lacked the combination of spectral range, single-photon purity, and reproducibility needed for a broadly useful NIR emitter platform (Figure \ref{fig: novelty}).}

\textcolor{black}{A variety of defect-engineering approaches have been explored in hBN to access new optical transitions, including plasma treatment, oxygen-related processing, irradiation, and thermal post-treatment (see Supplementary Note 1). These studies have established that plasma and oxygen can play an important role in modifying the defect landscape of hBN, but they have also shown that seemingly similar fabrication routes can lead to markedly different emitter populations and proposed microscopic origins. For example, plasma-treated hBN was shown to host single-photon emission \cite{xu2018single}, irradiation-assisted oxygen processing produced additional luminescent centers \cite{fischer2021controlled}, and more recent oxygen-related processing demonstrated narrowband emission associated with oxygen-involved color centers \cite{mohajerani2024narrowband, whitefield2026narrowband, tan2019robust}. Related oxygen-activated processing has also recently been shown to generate narrowband spin-addressable emitters \cite{whitefield2026narrowband}. Taken together, these studies make clear that oxygen/plasma processing in hBN is not associated with a single universal emitter class. Rather, modest changes in plasma conditions, annealing environment, excitation conditions, and local defect chemistry can stabilize distinct optical centers with different spectral windows, densities, and photophysical behavior (see Supplementary Note 1). This broader context is important as small variations in processing conditions can give rise to fundamentally different quantum-emitter platforms.}

\begin{figure*}[t!] \centering
     \includegraphics[scale=0.75]{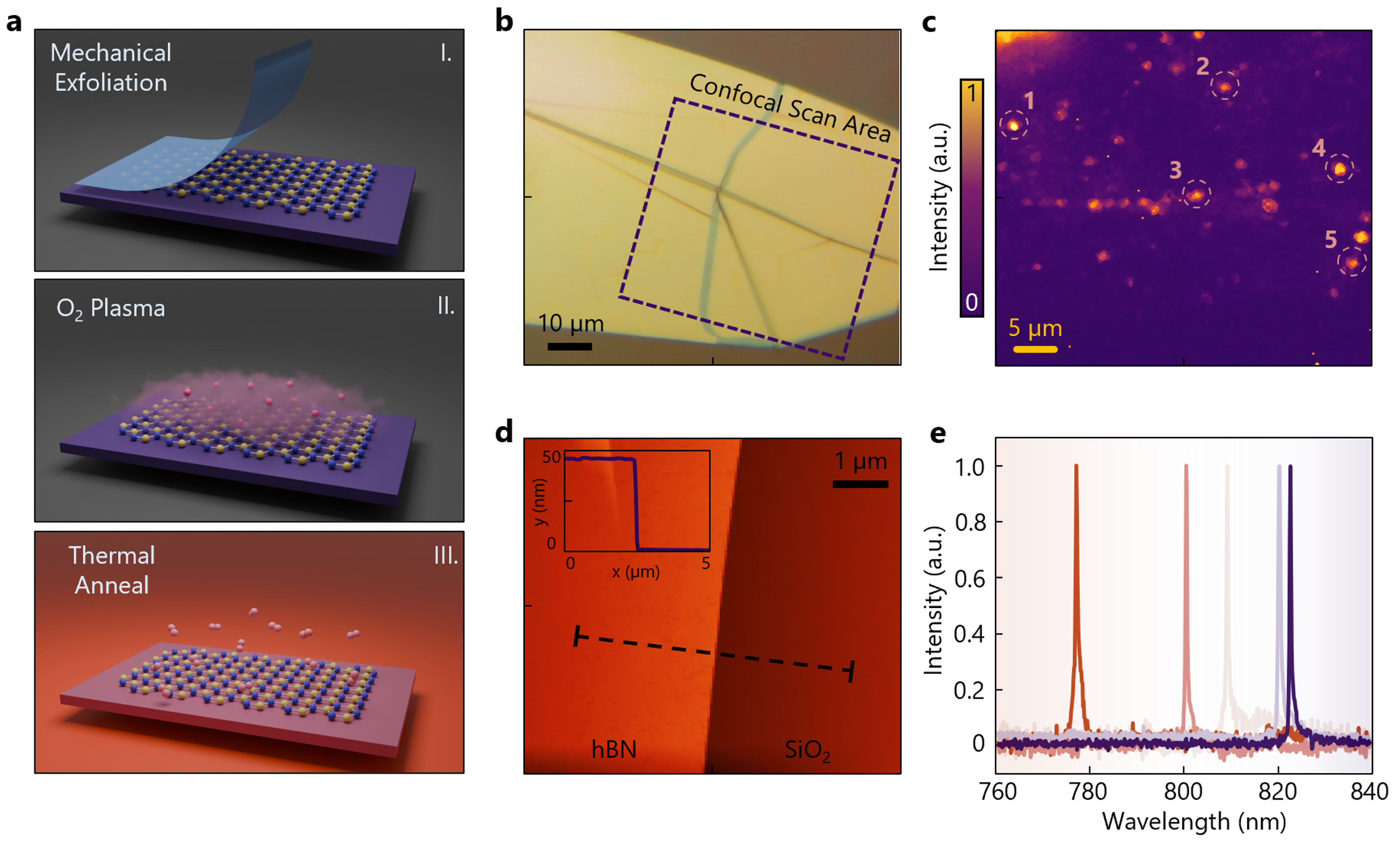}  
     \vspace{-0.5cm}
          \caption{\footnotesize \textbf{Fabrication process and overview of generated NIR SQEs.} \textbf{(a)} Schematic illustration of the fabrication process for generating NIR SQEs. Panel I denotes a mechanical exfoliation method to create sub-100-nm thick hBN samples. Panel II denotes an oxygen plasma treatment to introduce oxygen-related defects into the hBN lattice. Panel III denotes a thermal anneal process to stabilize the SQE within the hBN. \textbf{(b)} An optical micrograph of a processed hBN sample. The dashed line indicates the area of a photoluminescence (PL) confocal map. \textcolor{black}{\textbf{(c)} A confocal hyperspectral PL map of the processed hBN sample showcasing spatially resolved single-emitters across the sample.} \textbf{(d)} An atomic force microscopy (AFM) image is shown, highlighting the smooth surface topography of the hBN remains intact after the plasma treatment and processing. The inset shows a line scan across the dashed line and indicates a sample thickness of 50 nm. \textbf{(e)} Representative PL spectra of the generated spatially isolated NIR SQEs. Spectra of SQEs 1-5 in panel (c) are shown from left to right, respectively. SQEs are observed over a wide range of wavelengths spanning from 700 nm to 971 nm.}
          \label{fig: intro}
\end{figure*}

\textcolor{black}{Within this broader defect-engineering landscape, an important open question has been whether these processing strategies can be harnessed to realize a high-performance NIR emitter platform for free-space quantum networking applications. Previous studies have shown that hBN can access NIR optical transitions under selected processing and excitation conditions, establishing the broader possibility of quantum emission beyond the visible \cite{mohajerani2024narrowband, xu2018single, whitefield2026narrowband, taleb2025ultrafast}. At the same time, these reports have generally left open a more demanding materials challenge: the realization of an hBN single quantum emitter (SQE) platform in which the majority of emitters are within the technologically relevant 800--1000 nm regime with high purity, narrow spectral linewidths, and high brightness needed for quantum  photonic applications. In prior studies, emitters beyond 800 nm appear sparsely within broader datasets that are dominated by visible-range centers, and the most detailed optical characterization is correspondingly concentrated on emitters at shorter wavelengths (see Fig. \ref{fig: novelty}a and Supplementary Note 1). As a result, the combination of strong NIR prevalence, reproducible generation, extensive single-emitter characterization, and consistently high optical performance has remained an outstanding challenge.}

\textcolor{black}{Here, we address this need by demonstrating an oxygen-plasma-based route that reproducibly generates and establishes a distinct class of high-performance NIR SQEs in hBN. The SQEs generated here occupy a high-yield NIR regime, with nearly half of the studied emitters residing above 800 nm and emission wavelengths extending towards 1~$\mu$m. At the same time, these emitters combine high single-photon purity up to $99.9\%$ with many emitters demonstrating purity $>95\%$, ultra-narrow cryogenic linewidths under quasi-resonant excitation down to 11.1 $\mu$eV (2.7 GHz) (see Fig. \ref{fig: novelty}b), weak vibronic coupling with Debye–Waller factors approaching 50$\%$, MHz-level brightness (see Fig. \ref{fig: novelty}b), and high absorption and emission polarization visibility. They also exhibit strong photostability, including blinking-free operation, resistance to photobleaching, and sub-nm spectral stability over long time scales. These performance benchmarks are realized together within a single hBN SQE platform and are supported by a comparatively large statistical survey. Collectively, the results define a simple and scalable fabrication framework that delivers broad NIR coverage, strong statistical prevalence, robust stability, and high quantum-optical quality.}

\vspace{-10pt}
\section*{\label{results}Results and Discussion}
\vspace{-5pt}

\noindent\textbf{Defect Metrology}. Our oxygen-plasma defect creation process is illustrated in Fig. \ref{fig: intro}. We first perform a mechanical exfoliation of hBN bulk crystals grown in an AA’ stacking configuration to generate atomically thin layers of hBN ranging from 15-100 nm thickness with a subsequent standard solvent cleaning procedure (see Methods section) to remove organic residue from the surface (Fig. \ref{fig: intro}a I). Next, an oxygen plasma treatment is performed on the hBN flakes under a low vacuum environment to introduce oxygen impurities and vacancies at the surface and within the flakes (Fig. \ref{fig: intro}a II). Finally, a high temperature rapid thermal anneal (RTA) is performed under a nitrogen gas ($N_2$) and forming gas (90\% $N_2$, 10\% $H_2$) environment at 1000 $^\circ$C for 20 minutes (see Methods section) to allow the thermal relaxation of strain in the hBN flakes from the mechanical exfoliation process while also promoting the diffusion of oxygen and other impurities into the hBN to generate thermodynamically stable defect complexes that feature NIR emission (Fig. \ref{fig: intro}a III). To rigorously characterize the creation of these NIR SQEs, we have performed this fabrication across many thin hBN flakes.

\begin{figure*}[t] \centering
     \includegraphics[scale=1.1]{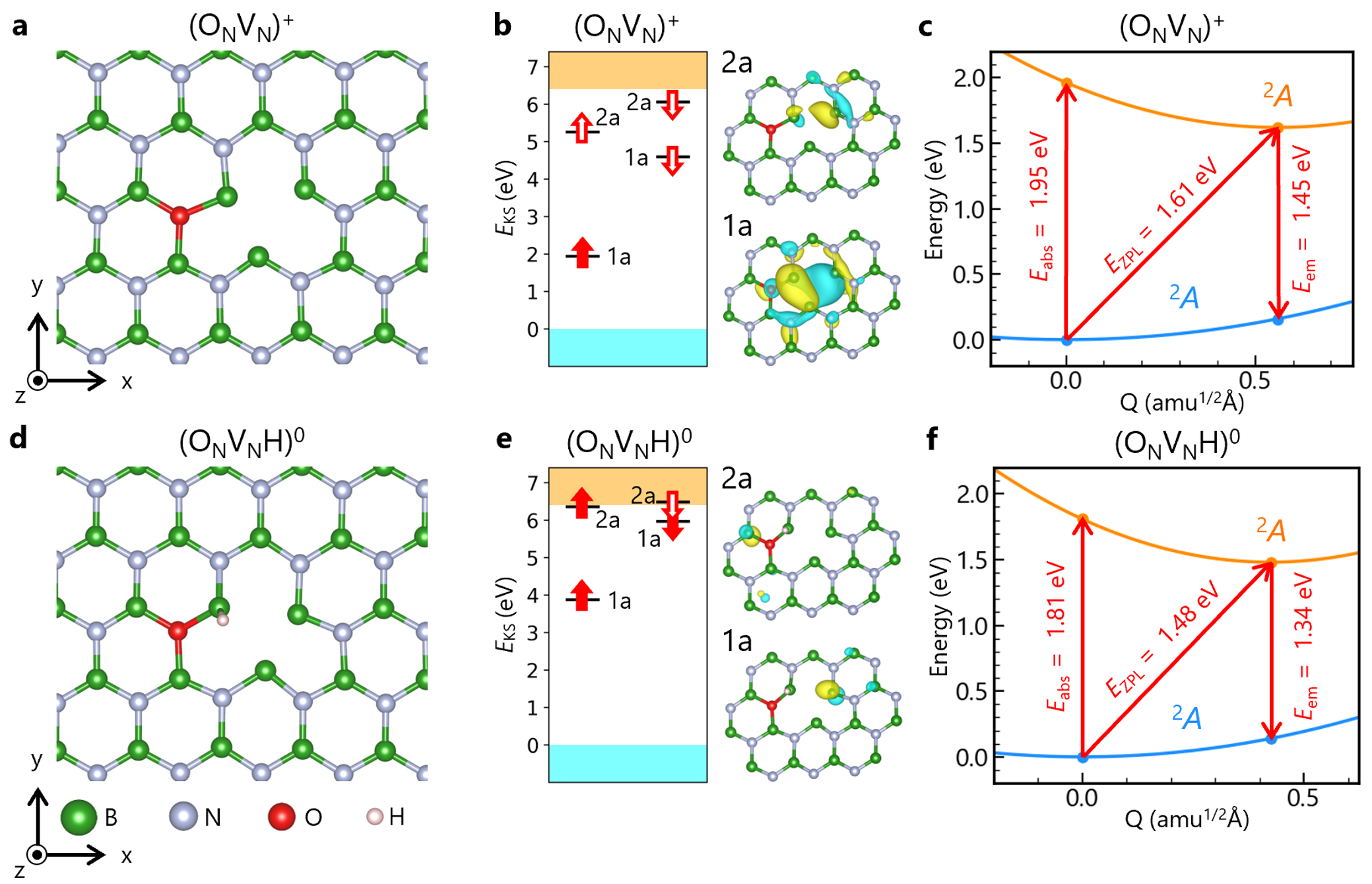}  
          \caption{\footnotesize \textbf{Properties of the O$_\mathrm{N}$V$_\mathrm{N}$ and the O$_\mathrm{N}$V$_\mathrm{N}$H centers.}
    (\textbf{a}) Structure of (O$_\mathrm{N}$V$_\mathrm{N}$)$^+$. 
    \textcolor{black}{(\textbf{b}) Left: Kohn-Sham states for the ground state $^2A$ of (O$_\mathrm{N}$V$_\mathrm{N}$)$^+$. Right: charge-density isosurfaces of the defect states of (O$_\mathrm{N}$V$_\mathrm{N}$)$^+$.} 
    (\textbf{c}) Calculated CCD for the $^2A \leftrightarrow~^2A$ transition.
    (\textbf{d}) Structure of (O$_\mathrm{N}$V$_\mathrm{N}$H)$^0$. 
    \textcolor{black}{(\textbf{e}) Left: Kohn-Sham states for the ground state $^2A$ of (O$_\mathrm{N}$V$_\mathrm{N}$H)$^0$. Right: charge-density isosurfaces of the defect states of (O$_\mathrm{N}$V$_\mathrm{N}$H)$^0$.} 
    (\textbf{f}) Calculated CCD for the $^2A \leftrightarrow~^2A$ transition.
    The isosurface absolute value is set to $4.0 \times 10^{-8}~e/a_0^3$.}
          \label{fig:theory}
\end{figure*}

Figure \ref{fig: intro}b displays an optical micrograph of a representative processed hBN flake of $\sim 50$~nm thickness undergoing the outlined fabrication procedure. The dashed box represents the hyperspectral confocal photoluminescence (PL) mapping scan area performed at 4~K, which is shown in Fig.~\ref{fig: intro}c. We observe high-density generation of localized hotspots corresponding to single-defect SQEs that emit in the NIR spectrum across a wide range of hBN flake thicknesses from 15--100~nm. Across 5 processed hBN flakes, we estimate an SQE area density of $2\times10^{6}$~cm$^{-2}$, indicating a high yield of SQE generation across a given hBN flake. \textcolor{black}{We note that we only observe strong NIR emission from such single defects with the inclusion of the oxygen-plasma treatment. Control experiments performed in the absence of the oxygen-plasma step, while maintaining the standard mechanical exfoliation and RTA procedure, show no NIR emission (Supplementary Note~3). In addition, EDS elemental mapping of representative flakes before and after plasma processing shows a clear post-treatment oxygen signal associated with the hBN region, indicating oxygen incorporation during fabrication (Supplementary Note~3). We emphasize that the EDS measurement is not intended to provide atomic-scale identification of the exact microscopic defect structure at the single-emitter level. Rather, it serves as a chemically informative correlation measurement and is most compelling when considered together with the reference-sample controls and the first-principles agreement with the proposed O$_N$V$_N$ and O$_N$V$_N$H defect candidates (see Fig. \ref{fig:theory}). We therefore interpret the combined evidence as strongly supporting oxygen incorporation as a necessary ingredient in the formation or activation of the observed NIR SQEs, while recognizing that the oxygen plasma may also assist defect creation indirectly through vacancy formation or related local structural modification.}

To analyze the surface topography of the processed hBN, we perform atomic force microscopy (AFM) and obtain AFM topography maps of the surface of multiple processed hBN flakes. Figure \ref{fig: intro}d shows a representative topography map of the processed hBN flake displayed in Fig. \ref{fig: intro}b with an atomic thickness shown in the step profile (inset, line cut taken along dashed line in Fig. \ref{fig: intro}d). Here, we extract an RMS surface roughness of 405.6 pm across the flake indicating the hBN surface remains relatively smooth and undamaged from the fabrication process allowing for generation of clean-interfaced 2D heterostructures and potential integration into optoelectronic and photonic devices. We observe a similar structural integrity and smooth surface topography across multiple samples used for this work. 

Furthermore, we see strong correlation between multiple hBN samples of these single defects being localized near hBN cracks, terraces, and folds that form prior to the plasma and RTA treatment. Such features are natural occurrences from the mechanical exfoliation process. This indicates a further correlation between these SQEs forming in areas of localized strain fields, irregular crystal symmetries, or a combination of both, which has been demonstrated in other types of defect complexes in hBN and other 2D materials \cite{mendelson2020strain,branny2017deterministic,parto2021defect}.\newline 

\noindent\textbf{Cryogenic Photophysics}. We next analyzed the spectral properties of the generated oxygen-related NIR SQEs. We obtain individual PL spectra from isolated single defects at 4 K indicated in Fig. \ref{fig: intro}e. Additional analysis on the room-temperature emission can be found in Supplementary Note 4. Notably, these SQEs exhibit sharp linewidths at cryogenic temperatures and span into the NIR wavelength range, making them highly suitable for free-space optical-based applications due to minimal atmospheric absorption losses in this spectral band \cite{jahid2022contemporary}. 

\begin{figure*}[t] \centering
     \includegraphics[scale=1.2]{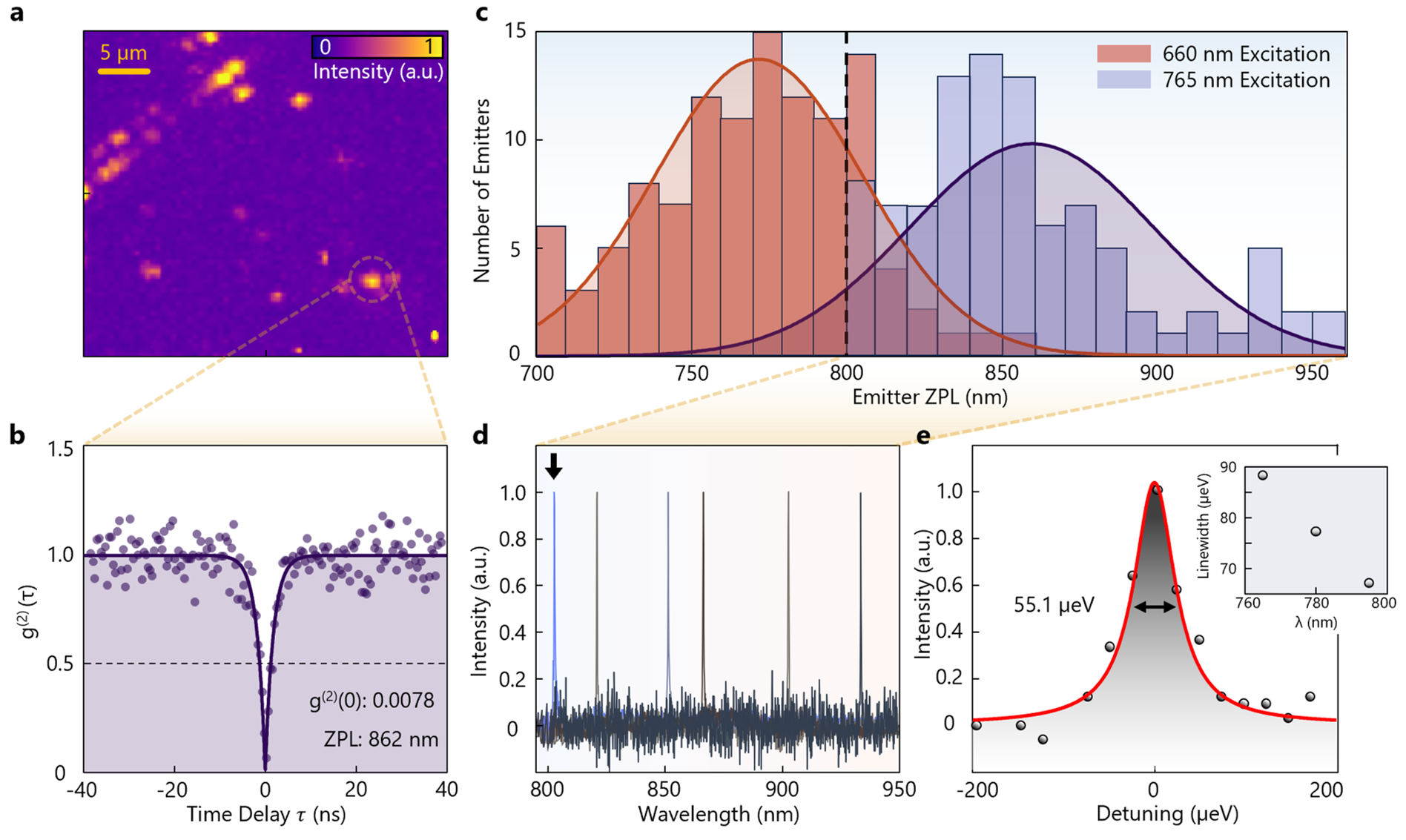}  
          \caption{\footnotesize \textbf{Characterization of NIR SQEs.} \textcolor{black}{\textbf{(a)} A confocal hyperspectral PL map at 4 K of a processed hBN sample. Local hotspots indicate spatially isolated NIR SQEs.} \textcolor{black}{\textbf{(b)} A select SQE is highlighted and indicated by the dashed circle in panel (a). The second-order autocorrelation function of the selected NIR SQE demonstrates photon anti-bunching with an uncorrected  $g^2(0)$ value of 0.0078 $\pm$ 0.0155.} \textbf{(c)} A histogram of spatially isolated single emitters from processed hBN sample in (a) and (b) displaying the span of ZPL wavelengths observed. The orange (blue) bins indicate the span of all isolated single emitters found under 660 nm (765 nm) pump excitation. A Gaussian fit indicates a center wavelength of 770 $\pm$ 33 and 856 $\pm$ 38 nm for 660 nm and 765 nm pump excitation, respectively. The dashed line represents the spectral location of a 800 nm longpass filter to reject the 765 nm pump laser. \textcolor{black}{The Gaussian fit for the 765 nm excitation subset only accounts for data on the right side of the dashed line.} \textbf{(d)} Representative PL spectra of spatially isolated NIR single emitters under 765 nm pump excitation. SQEs display spectral inhomogeneity spanning over a range of wavelengths spanning from 800 nm to 971 nm. \textbf{(e)} Quasi-resonant excitation of an NIR SQE with a ZPL of 801 nm indicated by arrow in (d) under 795 nm pump. A spectrometer-limited FWHM of 55.1 $\mu$eV (13.3 GHz) under pump power of 148 $\mu$W is extracted. The inset displays the linewidth dependence of the SQE on laser excitation wavelength from 765 nm to 795 nm under 400 $\mu$W pump power. }
          \label{fig: cryo characterization}
\end{figure*}

To verify the antibunching nature of these single-defect SQEs, we perform Hanbury Brown and Twiss (HBT) interferometry. We record the second-order photon correlation function $g^2(\tau)$ obtained under continuous laser (CW) excitation at 765 nm. \textcolor{black}{The result is presented in Fig. \ref{fig: cryo characterization}a,b of a representative SQE, which demonstrates a single photon antibunching dip at zero delay time with an uncorrected value of 0.0078 $\pm$ 0.0155 at zero time delay (99.22 $\pm$ 1.55 $\%$ single-photon purity), indicating  near-ideal anti-bunching from our NIR SQEs (see Anti-Bunching Dynamics Section for further details).}

\textcolor{black}{To investigate the stability of these NIR SQEs, we performed expanded photostability measurements spanning both long-timescale spectral monitoring and high-temporal-resolution intensity tracking (Supplementary Note 5). As shown in Fig.~S7a,b, a representative emitter measured on superconducting nanowire single-photon detectors (SNSPDs) remains stable over repeated measurements collected across three days, with a counts histogram that is well described by near-Poissonian statistics and yields a Fano factor of 1.097. This value being close to unity indicates that the intensity fluctuations remain close to the shot-noise limit, with no appreciable excess noise from intermittency or other non-Poissonian processes that would be expected for blinking or unstable emission. In addition, the spectral stability maps in Fig.~S7c--e show minimal to no measurable spectral wandering over one-hour measurements with 1~s integration and over shorter-timescale measurements with 100~ms integration, indicating blinking-free emission together with strong resistance to photobleaching. Across the expanded photostability datasets, we find that roughly 80\% of the NIR SQEs exhibit this level of stability, with no blinking and minimal-to-no spectral diffusion over the measured timescales, while the remaining emitters show only residual spectral wandering. For this minority subset, the diffusion is further suppressed at lower excitation powers, consistent with a reduction in pump-induced charge noise and local heating effects (see Fig. S8a,b). To reduce this slow-time scale diffusion, the photostability can be enhanced through further defect engineering methods such as surface passivation and active stabilization techniques \cite{li2017nonmagnetic,akbari2021temperature}.  Taken together, these results show that excellent photostability is a prevalent feature of this NIR emitter platform and supports its suitability for quantum-optical operation.}

\textcolor{black}{Next, we measured the integrated PL intensity as a function of excitation power to assess the brightness and saturable response of these NIR SQEs (Supplementary Note~5). In each case, the emission exhibits a clear saturation behavior, which is the characteristic signature of an individual localized optical transition with a finite excited-state occupation. We fit the data to a first-order saturation model \cite{allen2012optical}, $I = I_{\mathrm{sat}}P/(P + P_{\mathrm{sat}})$, where $P_{\mathrm{sat}}$ and $I_{\mathrm{sat}}$ are the saturation power and saturation intensity, respectively. As shown in Fig.~S6c, three representative NIR SQEs exhibit saturation powers of 704~$\mu$W, 3.0~mW, and 1.0~mW, with corresponding collection-corrected saturation intensities of 2.8~MHz, 5.8~MHz, and 1.9~MHz, respectively. These values demonstrate that the NIR defects are not only clearly saturable single emitters, but also bright, reaching emitted photon rates in the multi-megahertz regime without any optical-cavity enhancement. The observation of several-MHz saturation intensities is particularly notable for hBN emitters in this spectral range, where achieving both strong single-photon character and high brightness has remained challenging. The spread in saturation power and maximum emitted count rate further highlights emitter-to-emitter variation in local environment, excitation efficiency, and optical coupling, while at the same time showing that high brightness is not limited to a single exceptional SQE. Taken together, these measurements establish that this oxygen-related NIR emitter platform combines robust quantum-emitter behavior with reproducibly high photon flux, placing these SQEs among the brightest hBN single emitters reported to date in the near-infrared without optical-cavity enhancement \cite{xu2021creating}. We then perform polarization-resolved PL measurements to characterize the SQE polarizability where we observe high visibilities for both the absorption and emission dipoles alongside a distinct misalignment between them (see Supplementary Note 6 for further details).}

The excited-state lifetime of another representative SQE was extracted by performing time-resolved PL dynamics of the SQE using a pulsed laser excitation at 660 nm, resulting in an SQE radiative recombination lifetime of 1.74 $\pm$ 0.02 ns (Supplementary Note 5). SQEs measured from different samples have similar excited-state lifetimes on the order of $\sim$ 1-2 ns (Supplementary Note 5).

For the analysis of the ZPL distribution, we locate over 200 NIR SQEs across 5 different hBN flakes, all having undergone the same process undergoing 660 nm pump excitation (Supplementary Note 5). Here, we observe a Gaussian distribution forming with an average ZPL emission of 782 nm $\pm$ 36 nm. We note that with this fabrication process, we observe a few occurrences of SQEs in the visible spectrum across 550 nm - 700 nm, which have been associated with a boron vacancy, nitrogen vacancy, or carbon-related defect complexes \cite{mendelson2021identifying,sajid2018defect}. However, we observe a high density of single defect quantum emission above 800 nm, reaching ZPL wavelengths up to 971 nm, which has not been previously reported. Similar to hBN SQEs in the visible spectrum, we suspect the large spectral inhomogeneity of these NIR SQEs to arise from each single defect experiencing a variation in its local environment such as variable local deformation potentials due to the high strain coupling of atomically thin hBN flakes.

\begin{figure*}[t] \centering
     \includegraphics[scale=1.15]{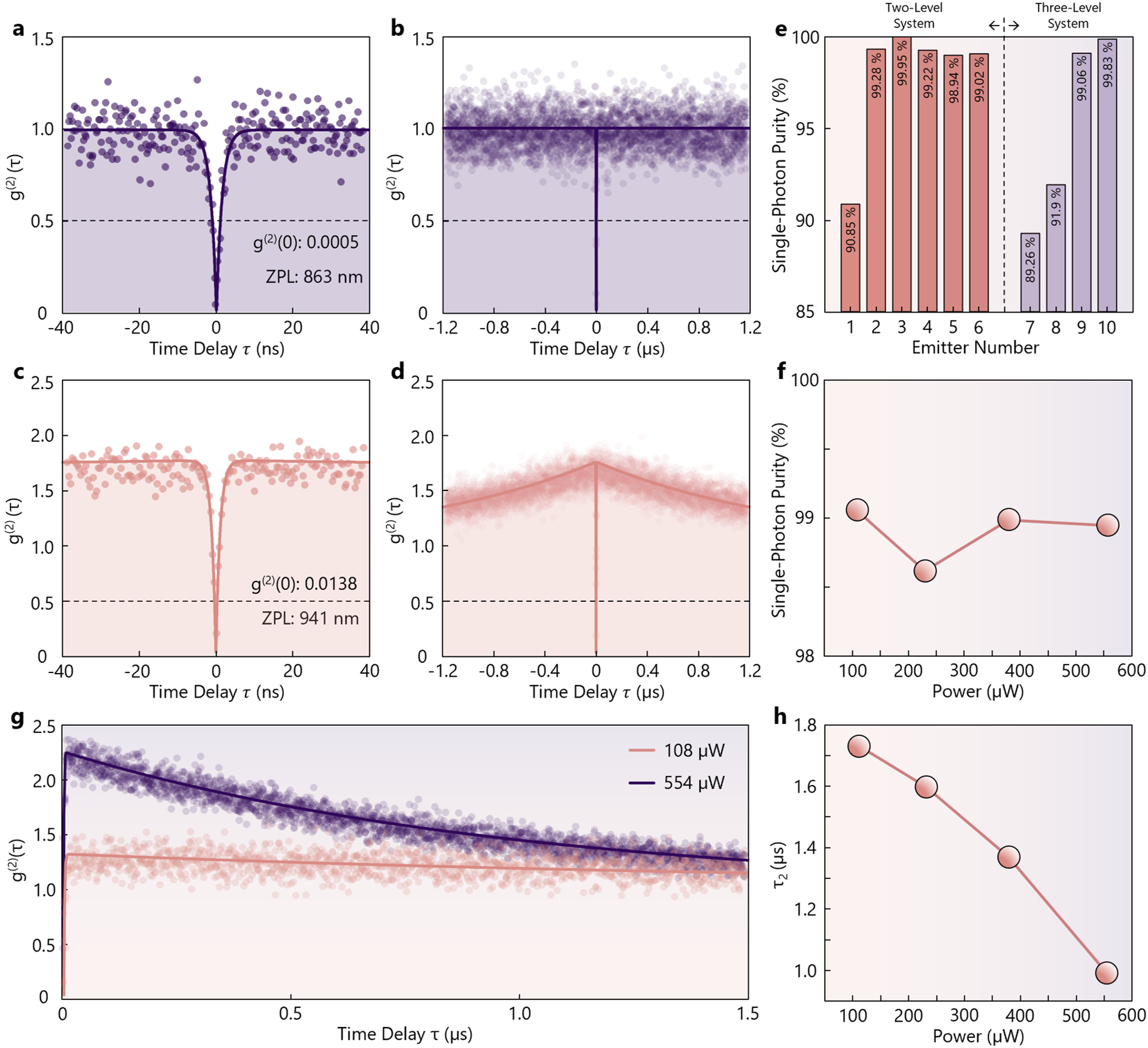}  
          \caption{\footnotesize \textbf{\textcolor{black}{Anti-bunching and bunching dynamics.}} \textcolor{black}{\textbf{(a)} Second-order autocorrelation function of a representative NIR SQE with ZPL at 863 nm, well described by a two-level system fit, exhibiting near-ideal antibunching with an uncorrected $g^2(0)$ of 0.0005 $\pm$ 0.0223. \textbf{(b)} Second-order autocorrelation function of the SQE in panel (a) measured over a longer time scale, showing no observable bunching dynamics. \textbf{(c)} Second-order autocorrelation function of a representative NIR SQE with ZPL at 941 nm, well described by a three-level system fit, exhibiting near-ideal anti-bunching with an uncorrected $g^2(0)$ of 0.0138 $\pm$ 0.0515. \textbf{(d)} Second-order autocorrelation function of the SQE in panel (c) measured over a longer time scale, showing pronounced bunching dynamics consistent with a three-level system containing a metastable shelving state. \textbf{(e)} Statistical summary of single-photon purity for 10 NIR SQEs, with the dashed divider separating emitters described by two-level and three-level $g^{(2)}(\tau)$ fits. \textbf{(f)} Single-photon purity versus excitation power for the SQE shown in (c) and (d), derived from the corresponding power-dependent $g^{(2)}(\tau)$ measurements. \textbf{(g)} Second-order autocorrelation function at positive delays for the SQE in (c) and (d) at 108 $\mu$W and 554 $\mu$W excitation power, showing increased bunching amplitude at higher pump power, consistent with power-dependent shelving dynamics in a metastable state. \textbf{(h)} Power dependence of the metastable-state lifetime $\tau_2$, obtained from the three-level autocorrelation analysis, showing that the effective shelving-state lifetime decreases with excitation power, indicative of pump-enhanced depopulation of the metastable state.}}
          \label{fig: g2}
\end{figure*}

We next examined the same emitters under optical pumping at excitation wavelengths closer to the average ZPL distribution, approaching quasi-resonant excitation utilizing a wavelength tunable 780 nm laser. Performing PL confocal mappings on a singular processed hBN flake, we obtain maps under 660 and 765 nm excitation of the same raster area as displayed in Supplementary Note 7. We observe the activation of NIR SQEs under 765 nm pump excitation that had previously remained in a dark state under 660 nm pump excitation. The distribution of all single emitters detected under 660 nm excitation (orange bins) for the hBN flake is shown in Supplementary Note 5, and an average ZPL emission of 770 $\pm$ 33 nm is extracted (see Fig. \ref{fig: cryo characterization}c). Collecting the same statistics under 765 nm excitation (blue bins), we observe an average ZPL emission of 856 $\pm$ 38 nm (see Fig. \ref{fig: cryo characterization}c). Our hypothesis, consistent with our first-principles modeling, is that the red shift in the average ZPL from 660 nm to 765 nm excitation is due to a “bright state” activation of SQEs emitting in the $>800$ nm spectral band that were previously “dark” or in lower densities under 660 nm excitation, as shown in Fig. \ref{fig: cryo characterization}c. \textcolor{black}{We also investigate the polarization dynamics under both pumping conditions, where we observe lower misalignment between the absorption and emission dipoles for the 765 nm excitation compared to 660 nm excitation (Supplementary Note 6).} We suspect this behavior arises from two related defects, O$_\mathrm{N}$V$_\mathrm{N}$ and O$_\mathrm{N}$V$_\mathrm{N}$H, supported by our modeling described in the following sections.

Next, we perform studies on the linewidth dependence on the excitation wavelength utilizing the same tunable 780 nm laser. Figure \ref{fig: cryo characterization}d highlights the representative spectral variety of single emitters at 4 K under 765 nm excitation collected from Fig. \ref{fig: cryo characterization}c. We observe a decrease in the linewidth of the SQEs under 765 nm excitation compared to previous SQEs displayed in Fig. \ref{fig: intro}e under 660 nm excitation. Fixing the excitation wavelength at 795 nm, we conduct a quasi-resonant excitation of a representative  single emitter with ZPL at 801 nm indicated by the arrow in Fig. \ref{fig: cryo characterization}d. We extract a spectrometer-limited linewidth of 55.1 $\mu$eV under a pump power of 150 $\mu$W, where correcting for the IRF yields an ultrasharp linewidth of 11.1 $\mu$eV (2.7 GHz). The inset displays a linewidth dependence of the same SQE on the excitation laser as it is tuned from 765 nm to 795 nm at a constant pump power of 400 $\mu$W. Here, we observe a decrease of the FWHM as the excitation wavelength approaches near resonance of the ZPL, which we attribute to a continuous suppression of spectral diffusion on fast time scales that arise from the sensitivity of the SQE to local charge fluctuations due to the DC-Stark effect. The fitted lineshape dependence on excitation wavelength from this SQE is shown in Supplementary Note 7. Earlier studies have already verified that spectral-diffusion can be strongly suppressed when hBN SQEs are driven either strictly at resonance \cite{tran2018resonant,konthasinghe2019rabi} or through anti-Stokes schemes \cite{tran2019suppression}. Here, we show that for these SQEs, linewidth reduction also occurs in the quasi-resonant ($<$ 40 meV detuning) range. 

\noindent\textbf{Anti-Bunching Dynamics.}
\textcolor{black}{To more comprehensively assess the single-photon purity and internal-state dynamics of these near-infrared SQEs, we performed an expanded set of HBT autocorrelation measurements under CW excitation at 765 nm. Figure~\ref{fig: g2}a shows the second-order autocorrelation function of a representative NIR SQE with a ZPL at 863~nm. The data are well described by the standard two-level model}

\begin{equation}
\textcolor{black}{g^{(2)}(\tau)=1-\alpha e^{-|\tau|/\tau_1},}
\end{equation}

\noindent\textcolor{black}{where $\alpha$ describes the antibunching contrast and $\tau_1$ is the characteristic radiative antibunching time. From the fit, we extract a near-ideal uncorrected value of $g^{(2)}(0)=0.0005\pm0.0223$, confirming high-purity single-photon emission. To verify that this emitter is well described by a simple two-level system and does not exhibit additional slow-state dynamics, we extended the correlation window to longer delay times. As shown in Fig.~\ref{fig: g2}b, no measurable bunching is observed on the microsecond timescale, indicating the absence of an appreciable metastable shelving pathway for this emitter within the measured range.}

\textcolor{black}{In contrast, a subset of the NIR SQEs exhibits more complex autocorrelation dynamics consistent with a three-level system. Figure~\ref{fig: g2}c shows a representative SQE with a ZPL at 941~nm that still displays pronounced antibunching, with an uncorrected value of $g^{(2)}(0)=0.0138$ $\pm$ $0.0515$, but whose full dynamics are better captured by the three-level form}

\begin{equation}
\textcolor{black}{g^{(2)}(\tau)=1-\alpha\left[(1+\beta)e^{-|\tau|/\tau_1}-\beta e^{-|\tau|/\tau_2}\right],}
\end{equation}

\noindent\textcolor{black}{where $\tau_1$ is the fast antibunching timescale associated with the optical cycle, $\tau_2$ is the metastable-state lifetime, and $\beta$ quantifies the bunching amplitude arising from population transfer into the shelving state. While the short-time trace in Fig.~\ref{fig: g2}c already establishes single-photon emission, extending the detection window to longer times in Fig.~\ref{fig: g2}d reveals clear bunching above unity. This behavior is consistent with a metastable shelving pathway \cite{fishman2023photon}, while also potentially reflecting other long-lived internal-state dynamics, such as charge-state switching or spin-dependent transitions, in which the emitter is intermittently diverted from the radiative cycle before returning to emission. Thus, the combined short- and long-timescale measurements distinguish two populations within these NIR SQEs, emitters that are well described by a two-level model and emitters that require a three-level description with a metastable state.}

\textcolor{black}{To evaluate how general this behavior is across the platform, we compiled the single-photon purity extracted from 10 distinct NIR SQEs, shown in Fig.~\ref{fig: g2}e. Of these, six emitters are well described by two-level behavior, while four exhibit clear three-level dynamics (see Supplementary Note 8). The vertical dashed line serves as a visual divider between these two classes. Across both groups, we repeatedly observe strong antibunching, with the best measured purities approaching 99.9\%, demonstrating that high-purity single-photon emission is not limited to a single exceptional emitter but is reproducibly achieved across this material platform. We stress that the distinction between the two groups is therefore not the presence or absence of single-photon emission, but rather whether additional long-timescale internal-state dynamics are present.}

\begin{figure*}[t!] \centering
     \includegraphics[scale=0.3]{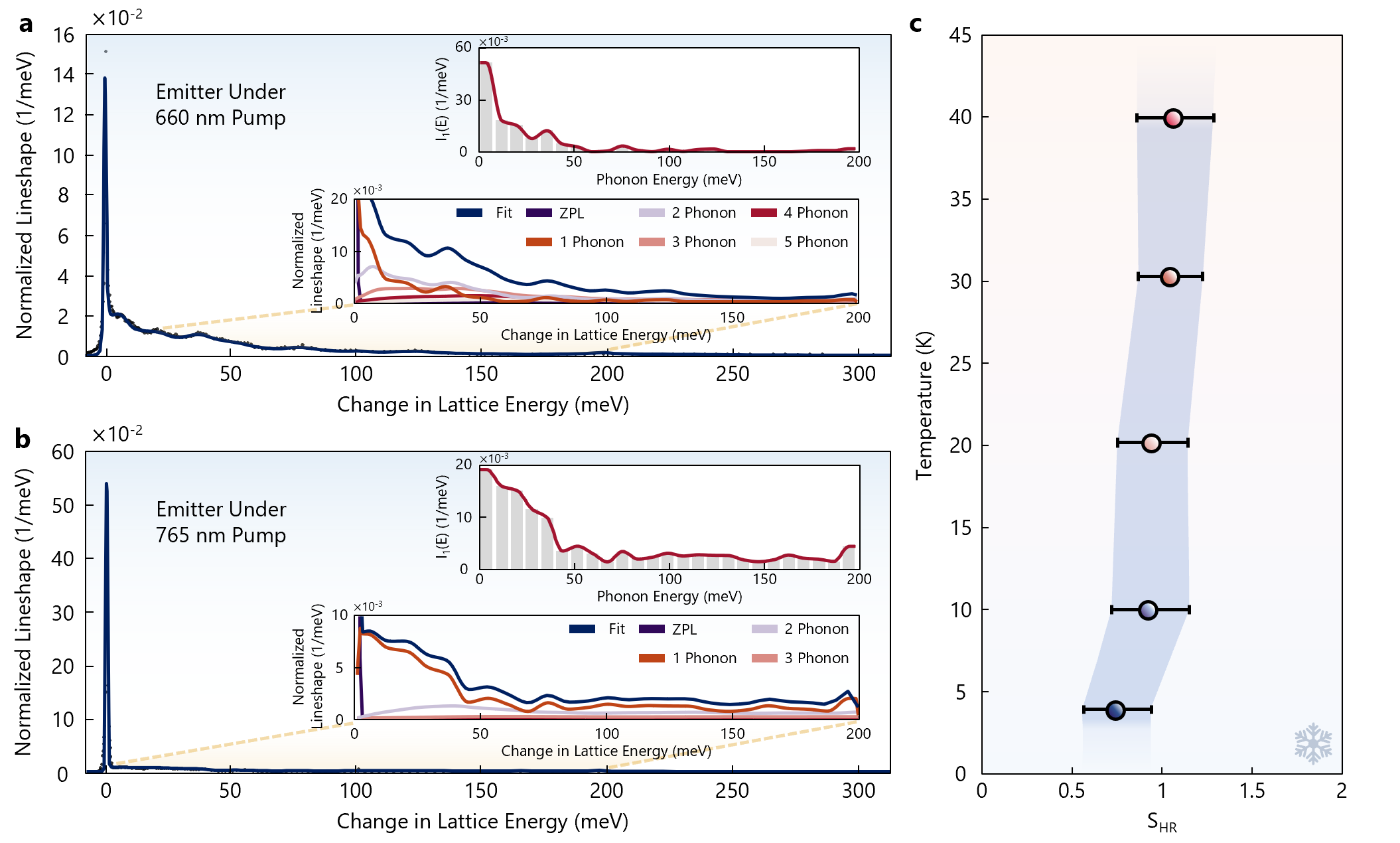}  
          \caption{\footnotesize \textbf{Electron-Phonon Coupling} \textbf{(a)} Normalized lineshape $L(\Delta  E)$ of an SQE excited under 660 nm pump with a fitted ZPL at 791.3 nm (1.567 eV). Bottom inset shows the zoom-in of the PSB with the $n$-phonon spectral distribution decomposition with an extracted $S_{HR}$ of 2.14 $\pm$ 0.40 \textcolor{black}{and DWF of 11.8 $\pm$ 5.7$\%$.} Top inset shows the 1-phonon vibronic coupling probability distribution function. \textbf{(b)} Normalized lineshape $L(\Delta E)$ of a SQE excited under 765 nm pump with a fitted ZPL at 859.9 nm (1.442 eV). Bottom inset shows the zoom-in of the PSB with the $n$-phonon spectral distribution decomposition with an extracted $S_{HR}$ of 0.72 $\pm$ 0.12 \textcolor{black}{and DWF of 48.7 $\pm$ 5.5$\%$}. Top inset shows the 1-phonon vibronic coupling probability distribution function. \textcolor{black}{\textbf{(c)} Temperature dependent Huang-Rhys analysis of the SQE under 765 nm pump in (b). Extracted Huang-Rhys factor, $S_{HR}$, at each temperature fall within one confidence interval of each other indicating temperature independent electron-phonon coupling strength over 4-40 K.} }
          \label{fig: EP coupling}
\end{figure*}

\textcolor{black}{We next examined how these dynamics evolve with optical pump power for the three-level emitter highlighted in Fig.~\ref{fig: g2}c,d. Figure~\ref{fig: g2}f shows that over the measured power range the extracted single-photon purity remains nearly unchanged and stays close to 99\%, indicating that increasing excitation power in this regime does not significantly degrade the antibunching at zero delay. This is important because it shows that the changes observed at longer delay times arise primarily from modified shelving-state kinetics rather than from a loss of single-photon character or the onset of appreciable multiphoton background.}

\textcolor{black}{The power dependence of the long-timescale autocorrelation is shown in Fig.~\ref{fig: g2}g for two representative excitation powers, 108~$\mu$W and 554~$\mu$W. At higher power, the bunching amplitude increases substantially, indicating that optical excitation more efficiently drives population into the metastable shelving state. This non-radiative or weakly emissive state leads to a stronger temporal clustering of emitted photons and therefore a more pronounced bunching signature. At higher optical power, the short-time single-photon purity and dynamics are unchanged, but the metastable state increases the probability that the emitter intermittently visits the shelving manifold.}

\textcolor{black}{At the same time, fitting the long-timescale autocorrelation traces yields the power-dependent metastable lifetime $\tau_2$, shown in Fig.~\ref{fig: g2}h. We observe that $\tau_2$ decreases monotonically with increasing excitation power, indicating that the effective residence time in the metastable state becomes shorter under stronger pumping. This behavior is consistent with pump-assisted depopulation of the shelving state \cite{fishman2023photon}, where optical excitation not only increases the rate at which population enters the metastable manifold, but also enhances the rate at which that population is returned to the radiative cycle. Panels~\ref{fig: g2}g and~\ref{fig: g2}h therefore support a common physical picture where with increasing pump power, the shelving state is populated more frequently, giving rise to a larger bunching amplitude, while simultaneously being emptied more rapidly, leading to a shorter extracted $\tau_2$. Together, these results show that a subset of the NIR SQEs possesses power-tunable three-level dynamics with a metastable shelving pathway, while still maintaining high single-photon purity. This rich internal-state structure may also be relevant for understanding the broader defect-level dynamics of these NIR SQEs. }

\noindent\textbf{Electron-Phonon Coupling}. The electron-phonon coupling of individual NIR SQEs is next analyzed using a finite-temperature Huang-Rhys vibronic model previously developed for hBN defects \cite{exarhos2017optical, patel2022probing}. In this framework, the PL spectrum of a defect is determined by the coupling between the electronic transition and lattice vibrations of the host crystal resulting in a sharp ZPL with PSBs on the low-energy side arising from multi-phonon emission processes. In the normalized PL lineshape fits, we extract the Huang-Rhys factor, $S_{\rm HR}$, which represents the average number of phonons involved in the emission event \textcolor{black}{and the Debye-Waller factor (DWF) which quantifies the fraction of emitted photons into the ZPL} using the free parameters of ZPL energy and width, and the discretized one-phonon coupling spectrum from which the full set of $I_n$ is constructed. A complete description of the spectral transformations, finite-temperature corrections, and fitting procedure is provided in the Methods section.

Figure \ref{fig: EP coupling}a shows the resulting fit for an SQE excited with 660 nm pump with a fitted ZPL of 791.3 nm (1.567 eV). The main panel displays the normalized lineshape $L(\Delta E)$ together with the best-fit decomposed into ZPL and multi-phonon contributions. The lower inset highlights the individual $n$-phonon components, while the upper inset shows the extracted one-phonon probability distribution $I_1(\Delta E)$. For this SQE, we obtain a Huang-Rhys factor $S_{\rm HR} = 2.14 \pm 0.4$ indicating that, on average, just over two phonons are involved in each emission event, which is consistent with relatively strong coupling to lattice vibrations. \textcolor{black}{This corresponds to a DWF of 11.8 $\pm$ 5.7$\%$, showing that only a modest fraction of the emitted photons are contained in the ZPL}. We then perform the same analysis on two additional SQEs excited at 660 nm to obtain an average Huang-Rhys factor $S_{\rm HR} = 1.70 \pm 0.16$ \textcolor{black}{and DWF of 18.3 $\pm$ 2.7$\%$}. The corresponding fits are provided in Supplementary Note 9. Fig. \ref{fig: EP coupling}b presents the same analysis for a second SQE excited with 765 nm pump with a fitted ZPL of 859.9 nm (1.442 eV). The overall lineshape and its phonon-resolved decomposition are again well described by the model, but the extracted Huang--Rhys factor is significantly smaller, $S_{\mathrm{HR}} = 0.72 \pm 0.12$. \textcolor{black}{This corresponds to a DWF of 48.7 $\pm$ 5.5$\%$, indicating that nearly half of the total emission is emitted into the zero-phonon line.} In the corresponding $I_1(\Delta E)$ panel, the features have lower probability densities and a modified spectral structure, consistent with a reduced overall strength of the one-phonon coupling and hence noticeably weaker vibronic coupling compared to the 660~nm pumped SQEs. \textcolor{black}{This relatively high Debye--Waller factor further highlights that the emitter operates in a substantially more ZPL-dominated regime.} Applying this procedure to two further SQEs excited at 765 nm yields an average Huang-Rhys factor $S_{\rm HR} = 1.04 \pm 0.10$ \textcolor{black}{and DWF of 35.3 $\pm$ 3.3$\%$,} with full fits shown in Supplementary Note 9. Taken together, these measurements show that SQEs addressed at 765 nm on average exhibit weaker vibronic coupling than those addressed at 660 nm.

In Fig. \ref{fig: EP coupling}c, we apply the same fitting procedure to temperature-dependent spectra of the 765 nm emitter from Fig. \ref{fig: EP coupling}b over the range 4--40~K. For each temperature we independently fit the full lineshape to extract $S_{\rm HR}(T)$. The resulting values cluster within a 3$\sigma$ confidence interval, showing no systematic trend with temperature. This behavior has several important implications. First, it confirms that the electron-phonon coupling is temperature independent in this range where the $S_{\rm HR}$ encodes the structural displacement between the ground- and excited-state lattice configurations, and the absence of any change in $S_{\rm HR}$ indicates that the defect does not undergo structural rearrangements, phase transitions, or ``softening'' as the sample is warmed from 4 to 40 K. Second, any observed evolution of the spectrum with temperature must therefore arise from changes in phonon occupation and from additional homogeneous broadening or dephasing, all of which are explicitly included in our model via the Bose-Einstein factors and the temperature-dependent ZPL width, further discussed in the next section. Third, the lack of a temperature dependence in $S_{\rm HR}$ over a range where $k_{\rm B} T $ is much smaller than typical optical-phonon energies is consistent with a picture in which the dominant vibronic coupling involves higher-energy phonon modes rather than very low-energy acoustic modes. Finally, the invariability of $S_{\rm HR}$ demonstrates that the intrinsic defect-phonon coupling strength is robust against modest temperature drifts, underscoring the stability of this emitter for cryogenic-temperature quantum optics applications.

\begin{figure*}[t] \centering
     \includegraphics[scale=0.37]{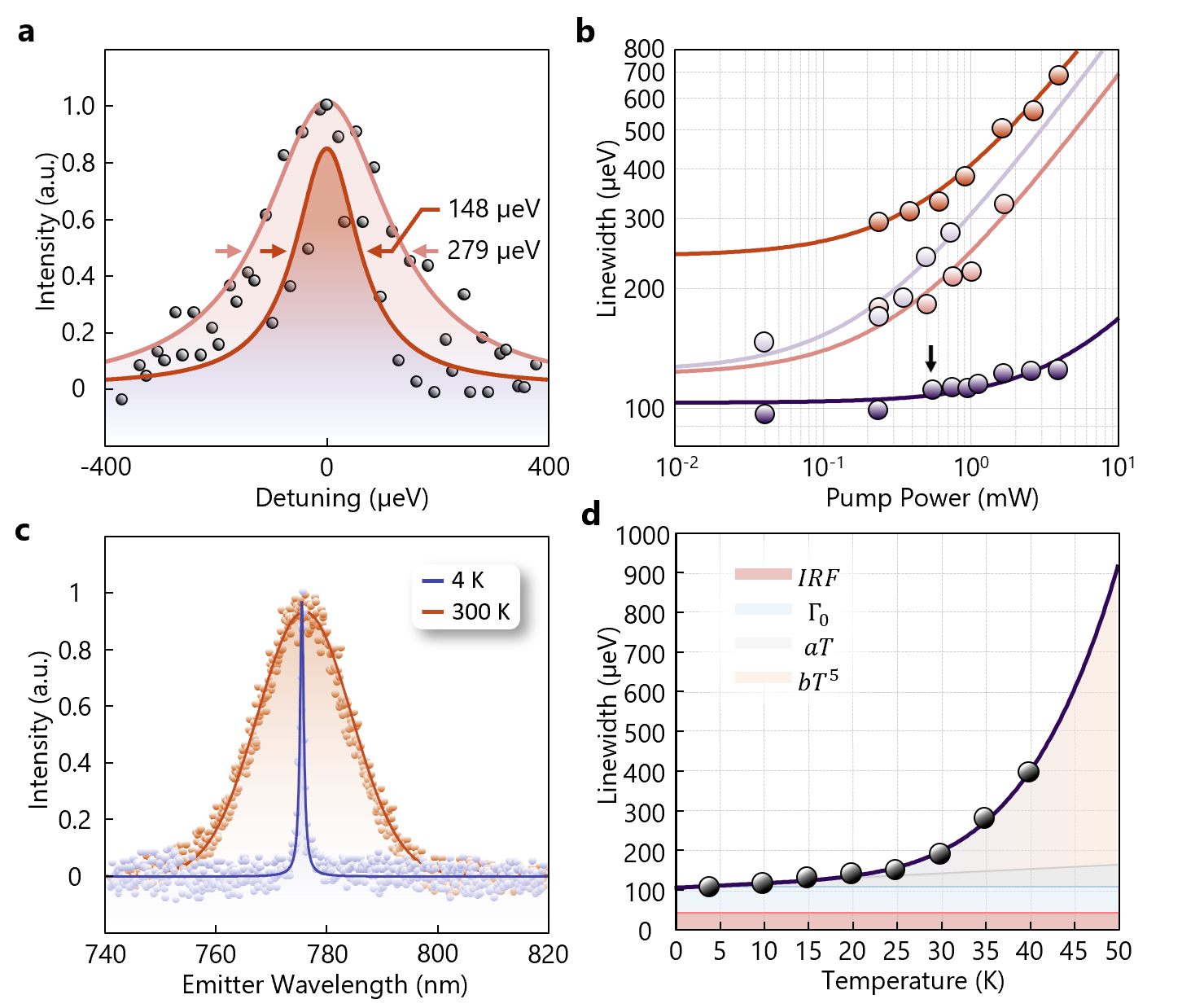}  
          \caption{\footnotesize \textbf{Non-resonant excitation linewidth characterization.} \textbf{(a)} A representative linewidth of an NIR SQE at 4 K under 40 $\mu W$ of pump power with a spectrometer-limited FWHM of 148 $\mu$eV compared to the same SQE under 3.4 mW of pump power with a spectrometer-limited FWHM of 279 $\mu$eV both under non-resonant pump excitation at 660 nm. \textbf{(b)} Linewidth power-broadening dependence of multiple NIR SQEs at 4 K. A two-level system power broadening model is fit to the experimental data points. \textbf{(c)} PL spectra of the representative SQE shown at both 4 K and 300 K. \textbf{(d)} Temperature-dependence of linewidth of a specific SQE taken under 500 $\mu W$ excitation power (indicated by arrow in panel (b)). The experimental data is fit to a model based on phonon-assisted broadening mechanisms coupled to deep and isolated electronic states of the defect. Shaded regions indicate contributions from different phonon-broadening mechanisms, instrument response, and homogeneous broadening. }  
          \label{fig: linewidth}
\end{figure*}

\noindent\textbf{Power and Temperature Broadening Dynamics}. With the sharp linewidths of our NIR SQEs, we next perform an extensive linewidth characterization at cryogenic temperatures. To extract the linewidth, we fit the experimental data with a Lorentzian function. In Fig. \ref{fig: linewidth}a, we show a representative NIR SQE at 4 K with a linewidth of 148 $\mu$eV (104 $\mu$eV under IRF correction) under low optical excitation power of 40 $\mu$W compared to power-broadend linewidth of 279 $\mu$eV (235 $\mu$eV under IRF correction) under high optical excitation power of 3.4 mW. 

We also examined pump power broadening of four NIR SQEs, shown in Fig. \ref{fig: linewidth}b. We fit the experimental data to a two-level pump power broadening system and extract the dephasing-limited zero-pump power linewidth, $\Gamma_0$, and the saturation pump power, $P_0$:

\vspace{-0.5cm}
\begin{equation}
    \Gamma = \Gamma_0 (1 + P/P_0)^{1/2}
    \label{eqn: power broadening}
\end{equation}

\noindent From our fits, we obtain uncorrected $\Gamma_0$ of 103.8, 122.6, 124.5, and 244.8 $\mu$eV from bottom to top respectively in Fig. \ref{fig: linewidth}b, corresponding to IRF corrected values of 59.8, 78.6, 80.5, and 200.8 $\mu$eV. The fitted lineshapes from these select SQEs are shown in Supplementary Note 7. These values are significantly lower than reported linewidth values for non-resonant excitation of typically several meV for SQEs that emit in the visible spectrum, which is a distinct feature of these NIR SQEs.

Figure \ref{fig: linewidth}c first shows the temperature dependent spectra at room temperature (300 K) and cryogenic temperature (4 K) of a representative NIR SQE with a ZPL at 775 nm which verifies the room-temperature operation under ambient conditions (Supplementary Note 4).

The temperature dependent linewidth of a representative NIR SQE is shown in Fig. \ref{fig: linewidth}d using an optical pump power of 500 $\mu$W. At the power indicated by the arrow in Fig. \ref{fig: linewidth}b, we measure the temperature dependence, shown in Fig. \ref{fig: linewidth}d), and fit the data to a least squares residual that follows the form: 

\vspace{-0.5cm}
\begin{equation}
    \Gamma(T) = \Gamma_0 + aT + bT^5
    \label{eqn: temp dep}
\end{equation}
\vspace{-0.5cm}

\noindent where we only consider phonon processes involving the linear and quadratic electron-phonon coupling terms which is valid for hBN due to its weak interactions with acoustic phonon processes \cite{ari2025temperature} as evident from the analysis in Fig. \ref{fig: EP coupling}. Here, the red shaded region in Fig. \ref{fig: linewidth}d represents the IRF. The blue shaded region represents the homogeneous broadening that sets the natural linewidth alongside the inclusion of inhomogeneous broadening mechanisms due to spectral diffusion on a fast time scale. These two regions contribute to the extracted zero-temperature linewidth $\Gamma_0$. As temperature initially increases, we observe the prominence of acoustic phonon processes from the piezoelectric coupling of hBN resulting in the linear dependence to $T$. Finally, as temperature continues to increase we observe deformation potential coupling which scales as $T^5$. This temperature dependence of the linewidth has been reported in other visible SQEs in hBN \cite{ari2025temperature} and other solid-state SQEs such as semiconductor quantum dots \cite{besombes2001acoustic} and defects in diamond \cite{jahnke2015electron}. For the SQE studied in Fig. \ref{fig: linewidth}d, we observe a narrow linewidth of 53.3 $\mu$eV correcting for the IRF at low pump power.

\noindent\textbf{First-Principles Calculations}. \textcolor{black}{We now employ first-principles calculations to help identify the most plausible structural and optical origins of the SQEs within the oxygen-related defect landscape suggested by the experiments. We focus our screening on oxygen-related defects and complexes because the NIR SQE population appears only when the O$_2$-plasma step is included, while EDS elemental mapping independently confirms oxygen incorporation after processing (see Supplementary Note 3). At the same time, we frame this assignment cautiously as these observations do not by themselves uniquely exclude all alternative plasma-enabled pathways, but they do strongly motivate an initial theoretical search within oxygen-containing defect families. This focus is further supported by prior literature on several alternative defect candidates, since a number of their reported or predicted ZPL energies do not align with the dominant NIR populations observed here, and in some cases their optical lifetimes are also inconsistent with experiment \cite{abdi2018color, li2020giant, sajid2020theoretical, xu2018single, dhu2024electrical, mendelson2021identifying, sajid2020vncb, li2022carbon}.}
We screen oxygen-related defects and complexes for ZPL energies and vibronic coupling.
Select configurations were already studied in \cite{weston2018native,li2022identification}.
The simplest candidate is substitutional oxygen on the nitrogen site (O$_\mathrm{N}$). 
We can exclude oxygen on the boron site (O$_\mathrm{B}$) because its formation energy is extremely large~\cite{weston2018native} and therefore is unlikely to exist as an isolated defect. 
O$_\mathrm{N}$ can form complexes with an oxygen interstitial (O$_\mathrm{N}$O$_{i}$, split interstitial), with a vacancy (O$_\mathrm{N}$V$_\mathrm{N}$, O$_\mathrm{N}$V$_\mathrm{B}$), or with carbon (O$_\mathrm{N}$C$_\mathrm{N}$ and O$_\mathrm{N}$C$_\mathrm{B}$), which commonly occurs as an unintentional impurity and has been linked to luminescence in hBN~\cite{mackoit2019carbon}. Since the sample consists of flakes that have edges or corners, we also investigate O$_\mathrm{N}$ in the vicinity of dangling bonds~\cite{turiansky2019dangling}, either in the same plane or in a neighboring plane. In addition, hydrogen is also a ubiquitous impurity in hBN~\cite{weston2018native} and can adsorb at structural imperfections such as cracks, terraces, and folds induced by mechanical exfoliation~\cite{turiansky2019dangling} or during the RTA process. Therefore, we also consider oxygen–hydrogen complexes, including O$_\mathrm{N}$H and O$_\mathrm{N}$V$_\mathrm{N}$H. In total, we investigated ten defect configurations, each in all possible charge states. Results are presented in Supplementary Note 10. Among them, complexes involving O\textsubscript{N}, nitrogen vacancies, and hydrogen emerge as the most likely candidates for the observed SQEs. In fact, the two sets of measured ZPLs can be correlated with two related defects: the O$_\mathrm{N}$V$_\mathrm{N}$ and O$_\mathrm{N}$V$_\mathrm{N}$H centers.

Our results for O$_\mathrm{N}$V$_\mathrm{N}$ are summarized in Fig.~\ref{fig:theory}. The ideal, unrelaxed structure would have $C_s$ symmetry. However, for all charge states the structures relax into lower-symmetry configurations.
In the positive charge state ((O$_\mathrm{N}$V$_\mathrm{N}$)$^+$, Fig.~\ref{fig:theory}a) the defect has $C_1$ symmetry and a spin-$1/2$ ground state denoted as the doublet $^2A$.
The B atoms near the vacancy display a small out-of-plane distortion of about 0.3~{\AA}. The Kohn–Sham level diagram of the $^2A$ ground state is shown in Fig.~\ref{fig:theory}b. 
Two states appear in the band gap. The \textcolor{black}{1a} state is a localized state formed by an antibonding combination of in-plane and out-of-plane $p$ orbitals on the B atoms, with a small contribution from the O $p$ orbital.
The \textcolor{black}{2a} state is mainly an in-plane $p$ orbital on one of the B atoms near the vacancy.
An internal excitation (calculated with $\Delta$SCF~\cite{jones1989density}) can promote an electron from \textcolor{black}{1a} to \textcolor{black}{2a} where the resulting configuration is also a $^2A$ state.

The configuration coordinate diagram (CCD) of the internal transition is shown in Fig.~\ref{fig:theory}c. The calculated ZPL energy of 1.62~eV (767~nm) agrees well with the SQEs observed under excitation at 660~nm (1.88~eV), which are clustered around 1.61~eV (770~nm) as shown in Fig.~\ref{fig: cryo characterization}c (orange bins). 
Our calculated Huang-Rhys factor, $S_{HR}=2.43$, matches well and is slightly larger than experimental values extracted in Fig. \ref{fig: EP coupling}a, but within computational and experimental uncertainties. 
A trend of computed HR factors being larger than values extracted from experiment has been observed in previous studies~\cite{razinkovas2021vibrational, turiansky_approximate_2025}. In addition, experimental Huang-Rhys factors tend to be underestimated due to the difficulty in distinguishing between zero-phonon photons and those coupled to low-energy acoustic phonons~\cite{jungwirth2016temperature}, and in identifying the full extent of the coupling to higher-energy phonons. 

The computed transition dipole moment for the transition is $\mu = 0.59$~e\AA{}, which yields a radiative lifetime of $\tau_{\mathrm{rad}} = 75$~ns. This value is longer than the experimentally measured PL decay time of $\sim 2$~ns for the 770~nm emitters, but these two quantities are not expected to coincide directly. \textcolor{black}{In the present calculations, $\tau_{\mathrm{rad}}$ corresponds to the intrinsic radiative lifetime of an idealized defect transition, whereas the experimentally measured lifetime ($\tau_{PL}$) is an effective PL decay time obtained under the actual excitation conditions and local sample environment, and therefore reflects both radiative $\tau_{rad}$ and nonradiative $\tau_{NR}$ processes:}
\begin{equation}
    \textcolor{black}{\tau_{\mathrm{PL}}=\frac{1}{1/\tau_{\mathrm{rad}}+1/\tau_{\mathrm{NR}}}=\eta\tau_{\mathrm{rad}}},
\end{equation}

\noindent\textcolor{black}{where $\eta$ is the quantum efficiency. A shorter measured PL lifetime is therefore physically reasonable for emitters with non-unity quantum efficiency, as is commonly the case in hBN \cite{li2022carbon}. In addition, the broader photophysical dataset indicates that some emitters exhibit internal-state dynamics beyond a simple isolated two-level transition, including metastable shelving behavior (see Fig. \ref{fig: g2}), which may further influence the experimentally observed decay kinetics. We therefore interpret the theory--experiment difference as arising from a combination of nonradiative relaxation, local defect-environment effects, and additional internal-state dynamics not fully captured in the idealized radiative-lifetime calculation.}

The equilibrium structure of the neutral (O$_\mathrm{N}$V$_\mathrm{N}$H)$^0$ complex is then shown in Fig.~\ref{fig:theory}(d). The defect is structurally similar to (O$_\mathrm{N}$V$_\mathrm{N}$)$^+$, except that it has an additional H atom bonded to a B atom between O$_\mathrm{N}$ and V$_\mathrm{N}$, with $C_1$ symmetry. (O$_\mathrm{N}$V$_\mathrm{N}$H)$^0$ is also a spin-1/2 center, with a doublet ground state $^2A$. The \textcolor{black}{1a} state (Fig. \ref{fig:theory}e) is a localized B $p_z$ state near the V$_\mathrm{N}$.
The higher \textcolor{black}{2a} state, being closer to the conduction band, exhibits more delocalized character (with a lower isosurface value) and includes contributions from the B $p_z$ orbital near O$_\mathrm{N}$. In the $^2A$ ground state, \textcolor{black}{1a} is doubly occupied and \textcolor{black}{2a} is singly occupied. 
When an electron is promoted from \textcolor{black}{1a} to \textcolor{black}{2a}, the excited $^2A$ state is formed.

The CCD of this transition is shown in Fig.~\ref{fig:theory}f. The calculated ZPL energy of 1.48~eV (840~nm) with a Huang-Rhys factor $S=1.73$ is close to the dominant emission of SQEs clustered around 1.45~eV (856~nm) that are observed under excitation at 765~nm (1.62~eV) (Fig. \ref{fig: cryo characterization}c, blue bins). The calculated Huang-Rhys factor once again matches well with the experimental observation shown in Fig. \ref{fig: EP coupling}b.

Our computed transition dipole moment for the (O$_\mathrm{N}$V$_\mathrm{N}$H)$^0$ transition is $\mu=0.51$ e{\AA}, resulting in a radiative lifetime $\tau_{\mathrm{rad}}=127$ ns. \textcolor{black}{Similar to the O$_N$--V$_N$ center, this calculated $\tau_{\mathrm{rad}}$ should be interpreted as the intrinsic radiative lifetime of an idealized defect transition and is therefore not expected to coincide directly with the experimentally measured PL decay time.} This longer lifetime (slower emission rate) may explain why the (O$_\mathrm{N}$V$_\mathrm{N}$H)$^0$ emitters are not prominently observed under shorter-wavelength (660 nm) excitation. The (O$_\mathrm{N}$V$_\mathrm{N}$H)$^0$ emitters are observed only when the excitation energy (765~nm) is low enough not to excite the (O$_\mathrm{N}$V$_\mathrm{N}$)$^+$ emitters. However, since our hyperspectral PL maps indicate that the two classes of emitters are not spatially co-localized, it is unlikely that they directly compete for pump power via their distinct lifetimes. Therefore, a more compelling explanation could be that the (O$_\mathrm{N}$V$_\mathrm{N}$)$^+$ emitters are excited more efficiently at 660 nm, while the (O$_\mathrm{N}$V$_\mathrm{N}$H)$^0$ emitters are preferentially excited at 765 nm, in line with the respective overlap of their vibronic excitation bands.

Our calculations are for O$_N$V$_N$ and O$_N$V$_N$H centers in ideal hBN crystals. The flakes in actual hBN samples lead to strains and distortions that can modify the properties of the SQEs. A study on similar defects in \cite{turiansky2021impact} has shown that realistic distortions can shift emission energies by several 0.1~eV, explaining the spectrum of ZPL wavelengths shown in Fig.~\ref{fig: cryo characterization}c. \textcolor{black}{In addition, our first-principles calculations indicate that both defect systems are optically active and host paramagnetic spin-doublet ground states, identifying them as potential candidates for ESR- and ODMR-active centers. We emphasize, however, that this spin-related assignment is based on theory alone and is not derived from direct experimental spin measurements. Rather, the calculated spin-doublet character suggests that these O$_N$V$_N$ and O$_N$V$_N$H centers may provide a pathway toward NIR spin-photon interfaces in hBN, which are naturally attractive for high-transmission free-space quantum networking. This therefore motivates future experimental studies aimed at directly probing the spin properties of these NIR SQEs, including ESR, ODMR, and optical spin-control measurements.}

\section*{Conclusion}
\textcolor{black}{In summary, we establish a distinct NIR SQE platform in hBN through a scalable oxygen-plasma treatment followed by rapid thermal annealing. Hyperspectral mappings of more than $>$300 emitters reveal a statistically significant NIR population spanning 700--971~nm, with nearly half of the studied SQEs occurring above 800~nm and strong reproducibility across processed flakes. Control samples further show that this NIR emitter population emerges only when the oxygen-plasma step is included, supporting a potential oxygen-related origin, while first-principles calculations identify O$_N$V$_N$ and O$_N$V$_N$H as the leading defect candidates with theorized paramagnetic spin-doublet ground states. Beyond establishing NIR emission in hBN, this work defines a distinct performance regime. Shown in Fig. \ref{fig: novelty}a, our platform achieves an order-of-magnitude higher fraction of color centers emitting above 800~nm than prior NIR hBN SQE reports while simultaneously extending the spectral reach by more than 100~nm toward the 1~$\mu$m regime at much higher emitter prevalence. Additionally, these emitters also occupy a favorable regime of simultaneously high brightness and ultranarrow linewidth, delivering higher first-lens brightness than prior NIR hBN emitters together with cryogenic linewidths in the 800+~nm band that are more than an order of magnitude sharper than previously reported NIR benchmarks (see Fig. \ref{fig: novelty}b). We further highlight that this platform combines these advantages with ultra-high single-photon purity, competitive areal density, and Debye--Waller values for hBN SQEs in the 800--1000~nm band approaching 50\% (see Supplementary Note 1).}

\textcolor{black}{The strength of this emitter class lies in the fact that these properties are realized together within one hBN platform. We observe single-photon purity approaching 99.9\%, excited-state lifetimes on the order of 1--2~ns, a high level of photostability with blinking-free operation, resistance to photobleaching, and sub-nm spectral stability over long time scales, as well as ultranarrow quasi-resonant linewidths down to 2.7~GHz. Complementary electron-phonon analysis shows that the 765~nm-addressed subset exhibits weaker vibronic coupling than the 660~nm subset, with temperature-independent Huang--Rhys behavior over 4--40~K, consistent with a comparatively clean and structurally robust phonon environment for coherent single-photon generation. To further assess transform-limited linewidths, resonant excitation schemes can be employed to minimize dephasing arising from above-band excitation and to suppress phonon-assisted processes, as has been demonstrated for other hBN quantum emitters \cite{tran2018resonant,dietrich2018observation,fournier2023investigating}. In parallel, the atomically thin nature of hBN can be leveraged to realize electrostatically gated vdW heterostructure devices in which the SQEs are embedded for dynamic charge-state control and carrier sweep-out, while also providing the most direct experimental route for disentangling oxygen incorporation from possible Fermi-level-assisted activation pathways in future studies. Here, our NIR SQEs are particularly well suited to such architectures, enabling more efficient suppression of spectral diffusion at lower gate voltages and deterministic coupling to nanophotonic cavities for Purcell enhancement and order-of-magnitude reductions in $T_1$, as demonstrated in our prior work \cite{parto2022cavity}. Moreover, the high observed dipole polarization visibilities together with a well-defined absorption and emission dipole offset provide a practical route to polarization-engineered and mode-matched emitter-to-cavity or emitter-to-waveguide interfaces. Taken together, these results position oxygen-related NIR SQEs in hBN as a scalable quantum-emitter platform that combines broad NIR coverage, strong statistical prevalence, robust stability, and high quantum-optical performance in a single vdW host material. This provides a practical pathway toward electrostatically stabilized and cavity-integrated NIR single-photon sources for free-space quantum communications, quantum networking and potential spin-photon architectures.}

\section*{Methods}
\subsection{Sample Preparation}
\textcolor{black}{We mechanically exfoliated hBN bulk single crystals grown in the AA’ stacking configuration onto a SiO$_2$/Si substrate. The SiO$_2$/Si substrate containing exfoliated hBN samples underwent a standard solvent cleaning procedure with 2 minute sonication in acetone followed by a 2 minute sonication in IPA followed by a dehydration bake at 115 $^\circ$C for 90 s. Next, the sample was placed inside a plasma asher (Technics PEII) undergoing a low frequency reactive ion etch (RIE) etch with O$_2$ flow undergoing plasma bombardment at low vacuum (300 mT) at 100 W for 1 minute. Finally, the samples undergo a rapid thermal anneal (AET RX6) at 1000 $^\circ$C for 20 minutes under a constant supply of N$_2$ and forming gas consisting of 90\% N$_2$ and 10\% H$_2$. The furnace was cooled down naturally under ambient conditions.}

\subsection{Optical Characterization}
A 660 nm continuous-wave PicoQuant laser diode and 780 tunable Toptica laser are used for steady-state PL measurements. A dichroic mirror at 680 nm cut-on and 70:30 beamsplitter is used to separate the optical excitation and collection paths for the 660 nm and 780 nm tunable laser, respectively. An additional 700 nm and 800 nm cut-on long-pass optical filter is used in the collection path to further extinguish the 660 nm and 780 nm tunable excitation laser, respectively, in the collection path. An infinity-corrected 0.42 NA NIR objective with 17 mm working distance is used for spectroscopy measurements. Our hBN samples are cooled to 4 K inside a Montana Instruments S200 cryostation and optical spectra are acquired using a Princeton Instruments HRS-500 spectrophotometer with 300/1200/1800 groove/mm gratings and a thermoelectrically-cooled Pixis Silicon CCD (see Supplementary Note 2 for further details). 

Second-order autocorrelation measurements with continuous-wave optical excitation are performed by utilizing the spectrophotometer as a monochromator to filter the emission from individual SQEs. The optical signal is then collected into a multimode fiber and HBT interferometry is performed using a broadband fiber-based beam splitter and two Excelitas SCPM-AQRH-13-FC single-photon avalanche detectors. Swabian Instruments time tagging electronics are used for photon counting and correlation. Furthermore, time-resolved PL measurements are performed using pulsed operation mode of the 660 nm laser diode with a 80 MHz repetition rate (see Supplementary Note 2 for further details). 

Polarization-resolved PL is conducted by modifying the steady-state PL setup with polarization optics. To measure the defect’s absorption dipole, a linear polarizer and half-wave plate are placed in the excitation path. The polarizer is set to maximize the optical power of the excitation laser, and the half-wave plate then rotates the input excitation beam. To measure the defect’s emission dipole, a linear polarizer is placed in the collection path and aligned to maximize the efficiency of our spectrophotometer. A half-wave plate is then placed prior to the polarizer and allows us to sweep through the defect’s emission spectrum (see Supplementary Note 2 for further details). 

\raggedbottom

\subsection{Finite-Temperature Phonon Sideband Analysis}
We analyze the phonon sidebands using a finite-temperature Huang-Rhys model adapted from \cite{exarhos2017optical, patel2022probing}. Measured PL spectra $I(\lambda)$ is collected (wavelength-dependent detection efficiency calibrated with broadband light source) and then re-binned to give a spectral distribution $S(\lambda)$. We convert to photon energy $E = hc/\lambda$ and define $S(E)$ such that $S(E)\,\mathrm{d}E = S(\lambda)\,\mathrm{d}\lambda$, which introduces the usual Jacobian factor. To remove the $E^{3}$ dependence of the radiative density of states, we work with the normalized lineshape
\begin{equation}
    L(E) = \frac{S(E)}{E^{3}},
\end{equation}
which isolates the intrinsic defect-phonon lineshape. In converting between these representations we propagate uncertainties from photon shot noise, detector background, and slow spectral drifts, and we adjust the energy binning to maintain approximately uniform error bars across the fitted range.

Following the Maradudin-Davies treatment of vibronic transitions in solids \cite{maradudin1966theoretical, davies1974vibronic}, we model $L(E)$ in terms of a ZPL and a PSB. The total spectrum as a function of lattice energy change $\Delta E = E_{\rm ZPL} - E$ is written as
\begin{equation}
    L(\Delta E) = e^{-S_{\rm HR}} I_0(\Delta E) + \big(I_0 \otimes I_{\rm PSB}\big)(\Delta E) ,
\end{equation}
where $S_{\rm HR}$ is the Huang-Rhys factor, $I_0(\Delta E)$ is the normalized ZPL lineshape, and $I_{\rm PSB}$ is the PSB in the limit of a delta-function ZPL where $\otimes$ denotes convolution. The PSB is expanded into contributions from processes involving $n$ phonons,
\begin{equation}
    I_{\rm PSB}(\Delta E) = \sum_{n} e^{-S_{\rm HR}} \frac{S_{\rm HR}^{n}}{n!}\, I_n(\Delta E),
\end{equation}
where the $I_n$ are normalized $n$-phonon probability distributions constructed recursively as $I_n = I_1 \otimes I_{n-1}$ for $n>1$. Finite temperature enters through the one-phonon distribution $I_1$, which we express in terms of an underlying phonon spectral function $S(E)$ and the Bose-Einstein occupation $n(E,T) = 1/(\exp(E/k_{\rm B}T) - 1)$ as
\begin{equation}
    I_1(E) =
    \begin{cases}
        A\,[n(E,T)+1]\,S(E), & E>0,\\[4pt]
        A\,n(|E|,T)\,S(|E|), & E<0,
    \end{cases}
\end{equation}
with $A$ chosen so that $\int I_1(E)\,\mathrm{d}E = 1$. Here $E>0$ corresponds to phonon emission and $E<0$ to phonon absorption, i.e., energy transferred from or to the lattice, respectively.

In practice, the phonon spectral function $S(E)$ is treated as an unknown function on the interval $0 \le E \le E_{\max}$, where $E_{\max}$ is taken to be the maximum relevant phonon energy in hBN ($\sim 200$~meV). We parameterize $S(E)$ by its values on a discrete energy grid $E_i = (i+\tfrac{1}{2})\,\delta E$ with step $\delta E$, and linearly interpolate between grid points. The free parameters in each fit are then the ZPL energy $E_{\rm ZPL}$, the ZPL width $\Gamma_{\rm ZPL}$, the Huang-Rhys factor $S_{\rm HR}$, and the set of discrete values $\{S(E_i)\}$. We perform weighted least-squares fits of the model $L(\Delta E)$ to the experimental lineshape, using the propagated uncertainties as weights. The resulting best-fit $S_{HR}$ and the derived one- and multi-phonon contributions $I_n(\Delta E)$ are reported in the main text and Supplementary Information.

\subsection{First-Principles Calculations}

The calculations are based on hybrid density functional theory with projector augmented wave (PAW) potentials~\cite{blochl1994projector}, as implemented in VASP~\cite{kresse1996efficient}. The plane-wave cutoff is set to 520 eV. We use the hybrid functional of Heyd, Scuseria, and Ernzerhof (HSE)~\cite{heyd2003hybrid,ge2006erratum}, combined with the Grimme-D3 correction for van der Waals interactions~\cite{grimme2010consistent}. The fraction of nonlocal Hartree–Fock exchange is set to $\alpha=0.40$. With this value, the calculated band gap is 6.40 eV, in agreement with the experimental gap~\cite{cassabois2016hexagonal} when zero-point renormalization from electron–phonon interactions~\cite{antonius2015dynamical} is included. The optimized lattice parameters are $a=2.48$ Å and $c=6.58$ Å, in good agreement with experiment ($a=2.50$ Å and $c=6.65$ Å~\cite{gu2007low}). This approach has been extensively tested, both in this work and in previous studies of h-BN~\cite{weston2018native,turiansky2019dangling}. Defects are modeled in a 240-atom conventional supercell with AA$^\prime$ stacking. The supercell is built from an orthorhombic cell ($a \times a\sqrt{3} \times c$) containing two primitive cells, scaled by $5\times3\times2$. Then the lattice vectors are fixed, while atomic positions around defects are relaxed until forces are below 0.01 eV/Å. Brillouin-zone sampling is performed using a single special k-point $(\tfrac{1}{4},\tfrac{1}{4},\tfrac{1}{4})$. Spin polarization is included in all calculations. 

To study internal transitions we employ the $\Delta$SCF methodology~\cite{jones1989density}. In this approach, excitation energies are obtained as total energy differences between two calculations with constrained occupations, each including full atomic relaxation. We use configuration coordinate diagrams (CCDs) to examine the coupling of these electronic transitions to lattice vibrations~\cite{dreyer2018first} and to compute Huang–Rhys (HR) factors $S$ within the one-dimensional approximation~\cite{alkauskas2012first}. The electron–phonon matrix elements are obtained from wavefunction overlaps and from linear-response theory, as implemented in the NONRAD code~\cite{turiansky2021nonrad}. 

We also study the radiative transition rate $\Gamma_{\mathrm{rad}}$, or equivalently the radiative lifetime $\tau_{\mathrm{rad}}$~\cite{stoneham2001theory,dreyer2020radiative}:
\begin{equation}
\Gamma_{\mathrm{rad}} = \frac{1}{\tau_{\mathrm{rad}}} = \frac{n_{D}E_{\mathrm{ZPL}}^{3}\,|\mu|^{2}}{3\pi \epsilon_{0}\hbar^{4}c^{3}},
\end{equation}
where $\epsilon_{0}$ is the vacuum permittivity and $n_{D}$ is the refractive index of h-BN. We use the experimental value $n_{D}=2.4$~\cite{cappellini2001optical}. $E_{\mathrm{ZPL}}$ and $\mu$ are evaluated explicitly from first principles.

\section*{Data Availability Statement}

\noindent The data that support the findings in this study are available from the corresponding author upon reasonable request.

\vspace{-10pt}
\section*{Supporting Information Available}
Details regarding \textcolor{black}{(1) Literature Benchmarking Tables, (2) Experimental Optical Setup and Polarization Calibration, (3) Chemical Analysis and Reference Samples, (4) Room Temperature Emission, (5) General Photophysics, (6) Polarization Resolved Analysis, (7) Linewidth Analysis, (8) Anti-Bunching Analysis,(9) Electron-Phonon Coupling Analysis, (10) First-Principles Calculations}


\begin{acknowledgments}
\noindent S.D., S.D.P., and G.M. thank Quantum Foundry, UCSB Nanofab, NRT at UCSB for valuable input and discussions. We gratefully acknowledge support from the NSF Quantum Foundry through Q-AMASE-i program Award No. DMR-1906325 and the UCSB NSF NRT Program Award No. 2152201. Y.C. and C.VdW. were supported by the U.S. Department of Energy, Office of Science, National Quantum Information Science Research Centers, Co-design Center for Quantum Advantage (C2QA) under contract number DE-SC0012704. M.E.T. was supported by the Office of Naval Research through the Naval Research Laboratory's Basic Research Program. J.A.G. and L.C.B. acknowledge support from the NSF Science and Technology Center for the Integration of Modern Optoelectronic Materials on Demand (award DMR-2019444). The research used resources of the National Energy Research Scientific Computing Center, a DOE Office of Science User Facility supported by the Office of Science of the U.S. Department of Energy under Contract No. DE-AC02-05CH11231 using NERSC award BES-ERCAP0021021.  
\end{acknowledgments}

\section*{Author Contributions}
\noindent S.D., and S.D.P. contributed equally to this work. G.M., S.D., and S.D.P. conceived the experiments. G.M. supervised the project. S.D. prepared the samples and performed the oxygen-plasma treatment, and thermal anneal. L.V. and L.J. assisted in the sample preparation. S.D. and S.D.P. performed steady-state PL spectroscopy, time-resolved PL spectroscopy, second-order auto-correlation, polarization-resolved spectroscopy, data processing, and formal analysis. Y.C., M.T., and C.VdW performed all first-principles calculations. L.C.B and J.A.G. performed the electron-phonon coupling data processing and analysis.

\section*{Competing Interests}
\noindent The authors declare no conflicts of interest.

\vspace{1 cm}
\bibliography{Biblio}
\end{document}